\documentclass[12pt]{article}
\usepackage{amssymb}
\usepackage[dvips]{epsfig}

\newcommand{\be}{\begin{equation}}
\newcommand{\ee}{\end{equation}}
\def\bea{\begin{eqnarray}}
\def\eea{\end{eqnarray}}

\newcommand{\bn}{\begin{eqnarray}}
\newcommand{\en}{\end{eqnarray}}

\newcommand{\barf}{\overline{f}}

\newcommand{\p}{\partial}

\newcommand{\nn}{\nonumber}
\def \ic {\'\i}

\newcommand{\no}{\noindent}

\newcommand{\tomega}{\tilde{\Omega}}

\newcommand{\s}{\,\,\,\,}
\def\bea{\begin{eqnarray}}
\def\eea{\end{eqnarray}}

\newcommand{\beq}{\begin{eqnarray}}
\newcommand{\eeq}{\end{eqnarray}}
\begin{document}

\title{\textbf{A note on linearized ``New Massive Gravity'' in arbitrary dimensions}}
\author{D. Dalmazi$^1$\footnote{dalmazi@feg.unesp.br} and R.C. Santos$^2$\footnote{renato.costa87@gmail.com}   \\
\textit{$^1${UNESP - Campus de Guaratinguet\'a - DFQ} }\\
\textit{{Avenida Doutor Ariberto Pereira da Cunha, 333} }\\
\textit{{CEP 12516-410 - Guaratinguet\'a - SP - Brazil.} }\\
\textit{$^2${ Instituto de F\ic sica Te\'orica - Universidade Estadual Paulista}} \\
\textit{{ Rua Doutor Bento Teobaldo Ferraz, 271, Bloco II}}\\
\textit{{ CEP 01140-070, S\~ao Paulo - SP, Brazil.}}}
\date{\today}
\maketitle

\begin{abstract}

By means of a triple master action we deduce here a linearized
version of the ``New Massive Gravity'' (NMG) in arbitrary
dimensions. The theory contains a 4th-order and a 2nd-order term
in derivatives. The 4th-order term is invariant under a
generalized Weyl symmetry. The action is formulated in terms of a
traceless $\eta^{\mu\nu}\Omega_{\mu\nu\rho}=0$ mixed symmetry
tensor $\Omega_{\mu\nu\rho}=-\Omega_{\mu\rho\nu}$ and corresponds
to the massive Fierz-Pauli action with the replacement
$e_{\mu\nu}=\p^{\rho}\Omega_{\mu\nu\rho}$. The linearized $3D$ and
$4D$ NMG theories are recovered via the invertible maps
$\Omega_{\mu\nu\rho} = \epsilon_{\nu\rho}^{\quad\beta}h_{\beta\mu}
$ and $\Omega_{\mu\nu\rho} = \epsilon_{\nu\rho}^{\quad
\gamma\delta}T_{[\gamma\delta]\mu} $ respectively. The properties
$h_{\mu\nu}=h_{\nu\mu}$ and $T_{[[\gamma\delta]\mu]}=0$ follow
from the traceless restriction. The equations of motion of the
linearized NMG theory can be written as zero ``curvature''
conditions $\p_{\nu}T_{\rho\mu} - \p_{\rho}T_{\nu\mu}=0$ in
arbitrary dimensions.

\end{abstract}

\newpage

\section{Introduction}

An important feature of general relativity is the infinite range
of the gravitational interaction mediated by a massless spin-2
particle. It is natural to speculate \cite{vdv,zak,db,vain}
whether the graviton might have a tiny mass which would certainly
have important consequences in the large scale physics of the
universe, see \cite{hinter} for a review work on massive gravity.

The use of the traditional Fierz-Pauli \cite{fp} theory for a
massive spin-2 particle leads \cite{vdv,zak} to a conflict, even
for a tiny mass, with experimental data for the deviations of
light beams by the sun which can be apparently solved by nonlinear
self-interaction terms \cite{vain}. The required nonlinearities
lead on their turn, in general, to the Boulware-Deser ghosts (lack
of unitarity) However, recent works, see \cite{rg,rgt} and also
\cite{hr12}, indicate that is possible to cope with unitarity by
fine tuning the nonlinear terms.

Another important issue in gravitational interactions is the lack
of renormalizability \cite{tv,gs}. The addition of
higher-derivative terms improve the UV behavior of the graviton
propagator but they bring up again the issue of ghosts
\cite{stelle}.

From the perspective of both IR and UV modifications of gravity
mentioned above, the 3D ``New Massive Gravity" (NMG) model
\cite{bht} plays an interesting role since it contains a massive
graviton and higher derivative (4th-order) term simultaneously
while keeping the theory unitary even beyond tree level
\cite{rgpty}. Moreover, the theory is invariant under general
coordinate transformations. All those nice features and its
relationship with $AdS_3/CFT_2$ duality \cite{sinha} have led to
several interesting works.

A step toward a generalization of \cite{bht} to $D > 3$ has been
taken in \cite{bfmrt} where a linearized version of a possible NMG
in $D=4$ is suggested. The model has been shown to be unitary.
While the 3D NMG model is usually formulated in terms of a
symmetric rank-2 tensor, the 4D model of \cite{bfmrt} is given in
terms of a rank-3 tensor satisfying $T_{[\mu\nu]\rho} =-
T_{[\nu\mu]\rho} $ and $T_{[[\mu\nu]\rho]}=0$.

Here we generalize the linearized version of the 3D  \cite{bht}
and 4D \cite{bfmrt} models to arbitrary dimensions by using a
rank-3 tensor which satisfies $\Omega_{\mu\nu\rho}= -
\Omega_{\mu\rho\nu} $ and is traceless
$\eta^{\mu\nu}\Omega_{\mu\nu\rho}=0$.

As an introduction to section IV, in sections II and III  we
derive 4th-order (in derivatives) higher-rank unitary descriptions
of spin-0 and spin-1 massive particles via the replacement $\phi
\sim \p_{\mu}B^{\mu}$ in the Klein-Gordon action and $A_{\mu} \sim
\p^{\nu}W_{\mu\nu}$, with $W_{\mu\nu}= W_{\nu\mu}$, in the
Maxwell-Proca theory. The use of a triple master action with
sources naturally explain why those simple change of variables do
not introduce ghosts.

Following closely sections II and III, in  section IV we show that
in the spin-2 case, a similar change of variables can be made in
the massive Fierz-Pauli theory formulated in terms of a
non-symmetric rank-2 tensor. We deduce the $D$-dimensional
linearized NMG model, see (\ref{lnmg}), and prove that it
correctly describes a massive ``spin-2'' particle in arbitrary
dimensions. We explain why the equations of motion of (\ref{lnmg})
can be written as a set of zero ``curvature'' conditions. In
section V we recover the linearized NMG in $D=3$ \cite{bht} and in
$D=4$ \cite{bfmrt}. In section VI we draw our conclusions and
perspectives.

\section{Higher derivative spin-0 model}

We can rewrite the Klein-Gordon action as a first-order theory by
lowering the order of the massless kinetic term via a vector
field:

\be S\left\lbrack \phi , A ,J \right\rbrack = \int \, d^D x \left(
\frac{m^2}2 A_{\mu}A^{\mu} + m\, A_{\mu}\p^{\mu}\phi - \frac{m^2}2
\phi^2 + J^{\mu}A_{\mu} \right) \quad . \label{sm1} \ee

\no  We have introduced an arbitrary source term for the vector
field for future purposes. Throughout this work we use
$\eta_{\mu\nu}={\rm diag}(-,+,\cdots , +)$.

If we integrate over the vector field in the corresponding
functional generating function we obtain

\be S\left\lbrack \phi , J \right\rbrack = \int \, d^D x \left(
-\frac 12 \p_{\mu}\phi \p^{\mu}\phi - \frac{m^2}2 \phi^2 -
\frac{J^{\mu}\p_{\mu}\phi}m - \frac{J_{\mu}J^{\mu}}{2\, m^2}
\right) \quad . \label{sphi} \ee

\no On the other hand, if we had first integrated over the scalar
field we would have obtained a higher-rank description of a spin-0
particle,

\be S\left\lbrack  A , J \right\rbrack = \int \, d^D x
\left\lbrack \frac 12 \left(\p_{\mu}A^{\mu}\right)^2 + \frac{m^2}2
A_{\mu}A^{\mu} + J^{\mu}A_{\mu} \right\rbrack \quad . \label{sa}
\ee

It can be shown that (\ref{sa}) correctly describes a spin-0
massive particle. However, the massless limit of the vector theory
does not describe a massless spin-0 particle as opposed to the
massless limit of (\ref{sphi}). The massless theory ${\cal
L}_{m=0}\equiv \left(\p_{\mu}A^{\mu}\right)^2/2 $ has no particle
content. It is invariant under gauge transformations

\be \delta_{\Lambda} A_{\mu} = \p^{\nu}\Lambda_{\mu\nu}
 \quad ; \quad \Lambda_{\mu\nu}= - \Lambda_{\nu\mu} \quad .
 \label{gt1} \ee

The equations of motion $\p_{\mu}\left(\p\cdot A\right) = 0 $
lead, with vanishing fields at infinity, to $\p \cdot A =0$, so we
are left with pure gauge degrees of freedom. It is convenient for
our purposes to check the trivial content of ${\cal L}_{m=0}$ by
lowering its order, see  (\ref{sm1}), , i.e.,   ${\cal L}_{m=0} =
- \phi^2/2 -\phi \, \p \cdot A $. The integral over the vector
field leads to the non-dynamic effective Lagrangian $-
\frac{\phi^2}2$ altogether with the functional delta function
$\delta\left( \p_{\mu}\phi \right)$ confirming the empty spectrum
of ${\cal L}_{m=0}$.

Since the massless term of (\ref{sa}) has no particle content, it
can be used as a ``mixing term'' in the master action approach of
\cite{dj} in order to derive another physically equivalent action
for a spin-0 particle. By introducing a dual vector field
$B_{\mu}$ we have:

\be S_M\left\lbrack  A,B,J \right\rbrack = \int \, d^D x
\left\lbrace \frac 12 \left(\p_{\mu}A^{\mu}\right)^2 + \frac{m^2}2
A_{\mu}A^{\mu} - \frac 12 \left\lbrack \p_{\mu}\left(B^{\mu} -
A^{\mu} \right)\right\rbrack^2  + J^{\mu}A_{\mu} \right\rbrace
\quad . \label{sm2} \ee

\no The shift $B_{\mu} \to B_{\mu} + A_{\mu} $ decouples the
vector fields and since the kinetic term for $B_{\mu}$ has no
propagating degree of freedom, it is clear that the particle
content of (\ref{sm2}) is the same of (\ref{sa}). On the other
hand, if we integrate over $A_{\mu}$ we have a fourth-order
description of a spin-0 particle:

\be S\left\lbrack  B,J \right\rbrack = \int \, d^D x \left\lbrack
\frac 1{2\, m^2} \p \cdot B \left(\Box - m^2\right) \p \cdot B
 + \frac{J^{\mu}\p_{\mu}\p\cdot B}{m^2} - \frac {J^{\mu}J_{\mu}}{2\, m^2} \right\rbrack \quad . \label{sb} \ee

 \no The action $S\left\lbrack  B \right\rbrack $ is invariant under the gauge transformation
  $\delta_{\Lambda} B_{\mu} $ given in (\ref{gt1}). Assuming vanishing fields at infinity, the equations
  of motion of (\ref{sb}), at vanishing sources, imply

  \be \left(\Box - m^2 \right) \p\cdot B = 0 \quad . \label{emb}
  \ee

  \no which reproduces the Klein-Gordon equation for the
  gauge invariant scalar $\p \cdot B$.

  Regarding unitarity of (\ref{sb}) we now look at the two-point amplitude saturated
   with sources. The $\Lambda$-gauge symmetry allows us to
  fix the gauge $F_{\mu\nu}(B)= \p_{\mu}B_{\nu} - \p_{\nu}B_{\mu}
  =0$. So we can add a gauge fixing term $-\lambda
  F_{\mu\nu}^2(B)$ to (\ref{sb}) and obtain the propagator. After the trivial
  redefinition $B_{\mu} \to \sqrt{2}\, m \, B_{\mu}$, the
  saturated two point function becomes

  \be A(k) = \tilde{J}_{\mu}^*(k) \left\langle A^{\mu}(-k)A^{\nu}(k)\right\rangle \tilde{J}_{\nu}(k) = \frac i2 \tilde{J}_{\mu}^* \left\lbrack
  \frac{\omega^{\mu\nu}}{k^2(k^2+m^2)} +
  \frac{\theta^{\mu\nu}}{\lambda\, k^2} \right\rbrack \tilde{J}_{\nu}
  \quad . \label{ak} \ee

  \no where $\tilde{J}_{\mu}$ is the source for the $B$-field and the spin-0 and spin-1 projection operators are given respectively
  by

  \be \omega_{\mu\nu} = \frac{k_{\mu}k_{\nu}}{k^2} \quad ; \quad
  \theta_{\mu\nu} = \eta_{\mu\nu} - \omega_{\mu\nu} \quad . \label{theta}
  \ee

The gauge invariance of the source term $\delta_{\Lambda} \int \,
d^D x \tilde{J}^{\mu}B_{\mu} =0$ requires longitudinal currents
$\tilde{J}_{\mu} (k) = i\, k_{\mu} J(k) $. Back in (\ref{ak}) we
have the gauge invariant amplitude for the $B$-model (\ref{sb}):

\be A(k) = \frac i2 \frac{\vert J \vert^2}{(k^2+m^2)} \quad .
\label{ak2} \ee

\no The imaginary part of the residue at $k^2=-m^2$ is positive
$R_{-m^2} = \vert J \vert^2/2 > 0 $. Therefore, we have one
massive spin-0 physical particle in the spectrum in agreement with
the original Klein-Gordon theory. Remarkably, after the
redefinition $B_{\mu} \to \sqrt{2}\, m \, B_{\mu}$, the massless
limit does also agree with the Klein-Gordon field theory. The
4th-order $B$-model corresponds to the original Klein-Gordon model
(\ref{sphi}) with the replacement $\phi \to - \left( \p \cdot B
\right)/m $ . In order to understand that point it is instructive
to define a triple master action

\be S_M\left\lbrack  A,B,\phi \right\rbrack = \int \, d^D x
\left\lbrace m\, A_{\mu}\p^{\mu}\phi - \frac{m^2}2 \phi^2 +
\frac{m^2}2 A_{\mu}A^{\mu} - \frac 12 \left\lbrack
\p_{\mu}\left(B^{\mu} - A^{\mu} \right)\right\rbrack^2  +
J^{\mu}A_{\mu} \right\rbrace \quad . \label{sm3} \ee

\no If we first shift $B_{\mu} \to \tilde{B}_{\mu} + A_{\mu} $ and
then integrate over $\tilde{B}_{\mu} $ and $A_{\mu}$ we obtain the
effective action (\ref{sphi}), whereas integrating over $\phi$
followed by the integral over $A_{\mu}$ gives rise to (\ref{sb}).
From derivatives with respect to the source $J^{\mu}$ for the
intermediate $A_{\mu}$ field we prove the equivalence of
correlation functions

\be \left\langle \p_{\mu_1} \phi (x_1) \cdots  \p_{\mu_N} \phi
(x_N) \right\rangle_{S[\phi,0]} = \frac 1{(-m)^N} \left\langle
\p_{\mu_1}\p\cdot B(x_1)  \cdots \p_{\mu_N}\p\cdot B(x_N)
\right\rangle_{S[B,0]} \label{cf} \ee

\no Since the quadratic terms in the sources in (\ref{sphi}) and
(\ref{sb}) are the same ones, the contact terms in (\ref{cf}) have
canceled out. So the correlation functions tend to match even at
coinciding points. So the dual map $\p_{\mu}\phi (x)
\leftrightarrow -\p_{\mu}\left[\p \cdot B(x)\right]/m $ holds
strongly and can be substituted inside the action where we have
coinciding points like $\p_{\mu}\phi (x) \p^{\mu}\phi(x)$. The
previous map and the boundary condition of vanishing fields at
infinity imply the map $\phi \leftrightarrow - \p \cdot B/m$. This
is to be compared to other dual maps between free theories, like
for instance, the map $f_{\mu} \leftrightarrow F_{\mu} \equiv
\epsilon_{\mu\nu\beta}\p^{\nu}A^{\beta}/m$ between the self-dual
model of \cite{pnt} and the Maxwell-Chern-Simons theory of
\cite{djt}. The correlation functions of $f_{\mu}$ only coincide
with correlation functions of $F_{\mu}$ up to contact terms, see
\cite{br}. If we replace $f_{\mu} = F_{\mu}$ directly in the
self-dual model we end up with a third-order theory with ghosts in
disagreement with Maxwell-Chern-Simons model of \cite{djt}.

\section{Higher derivative spin-1 model}

Analogously to the last section, we first rewrite the Maxwell-Proca theory in a first-order formalism by
lowering the order of the Maxwell theory with help of a symmetric tensor $W_{\mu\nu}=W_{\nu\mu}$ following the
appendix of \cite{kmu},

\be S\left\lbrack A, W,T \right\rbrack = \int \, d^D x \left\lbrack \frac{m^2}2 \left( W_{\mu\nu}W^{\mu\nu} -
\frac{W^2}{D-1}\right) + \sqrt{2} m \, W^{\mu\nu}\p_{\mu}A_{\nu} - \frac{m^2}2 A_{\mu}A^{\mu} +
W_{\mu\nu}T^{\mu\nu} \right\rbrack \quad . \label{sm4} \ee

\no where $T^{\mu\nu}=T^{\nu\mu}$ is the source.

Although the interacting term with the $W$-field depends only on the symmetric combination
$\p_{(\mu}A_{\nu)}=\left(\p_{\mu}A_{\nu} + \p_{\nu}A_{\mu}\right)/2$, if we integrate in the path integral over
$W_{\mu\nu}$ we get the Maxwell-Proca theory

\be S_{MP}\left\lbrack  A,T \right\rbrack = \int \, d^D x \left\lbrack -\frac 14 F_{\mu\nu}^2 - \frac{m^2}2
A_{\mu}A^{\mu} + \frac{\sqrt{2}}m T^{\mu\nu}\left(\eta_{\mu\nu}\p \cdot A - \p_{(\mu}A_{\nu)}\right) +
\frac{T^2-T_{\mu\nu}^2}{2\, m^2} \right\rbrack \quad . \label{smp} \ee

\no where $T=\eta_{\mu\nu}T^{\mu\nu}$.

On the other hand, starting with (\ref{sm4}) and integrating over
the vector field $A_{\mu}$ we have:

\be S\left\lbrack  W,T \right\rbrack = \int \, d^D x \left\lbrack  \left(\p^{\mu}W_{\mu\nu}\right)^2 +
\frac{m^2}2 \left( W_{\mu\nu}W^{\mu\nu} - \frac{W^2}{D-1}\right) + W_{\mu\nu}T^{\mu\nu} \right\rbrack \quad .
\label{sw} \ee

\no In \cite{ds} we have shown that (\ref{sw}) describes a massive spin-1 particle at classical and quantum
level and that the massless limit, similar to (\ref{sa}), is singular. The Lagrangian ${\cal L}_{m=0} =
(\p^{\mu}W_{\mu\nu})^2 $ has no particle content. The easiest way to check it is to lower its order rewriting it
as ${\cal L}_{m=0} = - A_{\mu}A^{\mu}/2 - A^{\mu}\p^{\nu}W_{\mu\nu} $. Integrating over $W_{\mu\nu}$ we have
$\p_{\mu}A_{\nu} + \p_{\nu}A_{\mu} = 0$ whose general solution, with vanishing fields at infinity, is trivial
$A_{\mu}=0$.

Based upon the similarities with the spin-0 case, the term $
(\p^{\mu}W_{\mu\nu})^2 $ can be used as a ``mixing term'' in a
triple master action with sources:

\bea S\left\lbrack A, W,H,T \right\rbrack &=& \int \, d^D x
\left\lbrace  \sqrt{2} m  \, W^{\mu\nu}\p_{\mu}A_{\nu} -
\frac{m^2}2 A_{\mu}A^{\mu} + \frac{m^2}2\left(
W_{\mu\nu}W^{\mu\nu} - \frac{W^2}{D-1}\right) \nn \right. \\
&-& \left.
 \left\lbrack \p^{\mu}\left(H_{\mu\nu} -
W_{\mu\nu}\right)\right\rbrack^2 + W_{\mu\nu}T^{\mu\nu}
\right\rbrace \, . \label{sm5} \eea

\no where we have introduced the symmetric dual field
$H_{\mu\nu}=H_{\nu\mu}$. On one hand, if we shift $H_{\mu\nu} \to
H_{\mu\nu} + W_{\mu\nu}$ and integrate over $H_{\mu\nu}$ and
$W_{\mu\nu}$ we derive the Maxwell-Proca theory given in
(\ref{smp}), whereas the functional integration over $A_{\mu}$
followed by the integral over $W_{\mu\nu}$ leads to the 4th-order
model:

\be S\left\lbrack H,T \right\rbrack  \!\! = \! \! \int d^D x
\left\lbrace - \frac 14 F_{\mu\nu}^2 [ \p H ] - \frac{m^2}2
\left(\p^{\nu}H_{\mu\nu}\right)^2  - \frac{\sqrt{2}
T^{\mu\nu}}{m}[\eta_{\mu\nu}\p^{\alpha}\p^{\beta}H_{\alpha\beta} -
\p^{\alpha}\p_{(\mu}H_{\nu)\alpha}] + \frac{T^2-T_{\mu\nu}^2}{2\,
m^2} \right\rbrace \, \label{sh} \ee

\no where $ F_{\mu\nu} [ \p H ] =
\p_{\mu}(\p^{\alpha}H_{\nu\alpha}) -
\p_{\nu}(\p^{\alpha}H_{\mu\alpha})$.

 As in the spin-0 case, the quadratic terms in the sources in
(\ref{smp}) and (\ref{sh}) are exactly the same
 which leads us to identify correlation functions of $\eta_{\mu\nu}\p \cdot A - \p_{(\mu}A_{\nu)}$
 in the Maxwell-Proca theory with correlation functions of $-\eta_{\mu\nu}\p^{\alpha}\p^{\beta}H_{\alpha\beta}
 + \p^{\alpha}\p_{(\mu}H_{\nu)\alpha}$ in the $S\left\lbrack H,0
\right\rbrack $ theory. The contact terms cancel out again.
Consequently, we have the strong dual map
  $A_{\mu} \leftrightarrow -\p^{\alpha}H_{\mu\alpha}$ which can be used inside Lagrangians. Thus, explaining the
  Maxwell-Proca form of (\ref{sh}). Notice also that $S\left\lbrack H,0 \right\rbrack $ is invariant under any
  local transformation  which preserves $\p^{\alpha}H_{\alpha\beta}$. They can be written \cite{dst} as
  $\delta_B H_{\mu\nu} = \p^{\sigma}\p^{\rho}B_{\mu\sigma\rho\nu} $ where the gauge parameters
  $B_{\mu\sigma\rho\nu}$ have the same index symmetries of the Rieman tensor.

   The equations of motion of (\ref{sh}) are

  \be \p_{\mu}V_{\nu} + \p_{\nu}V_{\mu} =0 \quad ; \quad {\rm where} \quad
  V_{\nu} \equiv \left\lbrack \eta_{\mu\nu}(\Box -m^2) - \p_{\mu}\p_{\nu}\right\rbrack \p_{\sigma}H^{\mu\sigma}
  \quad . \label{eqm2} \ee

\no With vanishing fields at infinity we have the general solution
$V_{\mu}=0$ which is equivalent to the Maxwell-Proca equations
with the identification of the vector field $A_{\mu}$ with the
gauge invariant $ \p^{\sigma}H_{\mu\sigma} $, which proves the
classical equivalence of the 4th-order model (\ref{sh}) with the
Maxwell-Proca theory.

Concerning unitarity, first it is convenient to introduce sources for $H_{\mu\nu}$ and  write (\ref{sh}) in
terms of spin projection operators:

\be S\left\lbrack H,T \right\rbrack  =  \int d^D x \left\lbrace H_{\lambda\mu}\left\lbrack -\frac{\Box (\Box -
m^2)}2 P_{SS}^{(1)} + m^2 \, \Box \, P_{WW}^{(0)}\right\rbrack^{\lambda\mu}_{\s\s\alpha\beta}H^{\alpha\beta} +
H_{\mu\nu}\tilde{T}^{\mu\nu} \right\rbrace \, \label{shtt} \ee

\no where the above spin-1 and spin-0 projection operators are given respectively by

\be \left( P_{SS}^{(1)} \right)^{\lambda\mu}_{\s\s\alpha\beta} = \frac 12 \left(
\theta_{\s\alpha}^{\lambda}\,\omega^{\mu}_{\s\beta} + \theta_{\s\alpha}^{\mu}\,\omega^{\lambda}_{\s\beta} +
\theta_{\s\beta}^{\lambda}\,\omega^{\mu}_{\s\alpha} + \theta_{\s\beta}^{\mu}\,\omega^{\lambda}_{\s\alpha}
 \right) \quad , \label{pss1} \ee

\be \left( P_{WW}^{(0)} \right)^{\lambda\mu}_{\s\s\alpha\beta} = \omega^{\lambda\mu}\omega_{\alpha\beta} \quad ,
\label{pww0} \ee

\no Secondly, we can add a gauge fixing term $\lambda \,
G_{\mu\nu}^2 (H)$ to the action, where

\be G_{\mu\nu}(H) \equiv \Box H_{\mu\nu} - 2
\p^{\alpha}\p_{(\mu}H_{\nu ) \alpha} + \eta_{\mu\nu}
 \p^{\alpha}\p^{\beta}H_{\alpha\beta} = 0 \quad , \label{gc} \ee

 \no defines a good gauge condition. It is symmetric and transverse $\p^{\mu}G_{\mu\nu} = 0$ just like the gauge parameter
 $\p^{\sigma}\p^{\rho}B_{\mu\sigma\rho\nu}$. Now we are able to obtain
 the propagator and the two-point amplitude saturated with
 sources. Since the gauge symmetry is such that $\delta_B\p^{\nu}H_{\mu\nu}=0$, the invariance of the source term $\delta_B
 \int d^D x H_{\mu\nu}\tilde{T}^{\mu\nu} =0$  requires  $\tilde{T}_{\mu\nu} = \p_{\mu}J_{\nu} +
 \p_{\nu}J_{\mu}$. Taking into account such restriction we have

 \bea A(k) &=& \tilde{T}_{\nu\alpha}^*\left\langle H^{\nu\alpha}(-k)H^{\sigma\mu}(k)\right\rangle
 \tilde{T}_{\mu\sigma}(k) = i\,
  \tilde{T}_{\nu\alpha}^*\left\lbrack \frac{2P_{SS}^{(1)}}{k^2(k^2+m^2)} +
 \frac{P_{WW}^{(0)}}{m^2 \, k^2} \right\rbrack^{\nu\alpha\mu\sigma}\tilde{T}_{\mu\sigma} \nn\\
&=& i \left(\frac{J^*_{\mu}(k)\theta^{\mu\nu}J_{\nu}(k)}{k^2 +
m^2} + \frac{J^*_{\mu}(k)\omega^{\mu\nu}J_{\nu}}{m^2}\right)
\label{akmp} \eea

\no where we have used $\tilde{T}_{\mu\nu} = i ( k_{\mu}J_{\nu} +
k_{\nu}J_{\mu})$. The last line of (\ref{akmp}) corresponds
exactly to the two-point amplitude of the Maxwell-Proca theory
with sources $J_{\mu}$. The pole at $k^2=0$ inside the operators
(\ref{theta}) cancel out and the residue at $k^2=-m^2$ is of
course positive, see for instance \cite{ds}, which guarantees
unitarity of our higher derivative Maxwell-Proca theory.

Before finishing this section we point out that we could have
started from a simpler first-order version of the Maxwell-Proca
theory where an antisymmetric field $B_{\mu\nu}=-B_{\nu\mu}$
(2-form) is introduced to lower the order Maxwell theory instead
of the symmetric field $W_{\mu\nu}$ of \cite{kmu} and followed the
same steps to arrive at a 4th-order theory. However, the 4th-order
model would not be ghost-free. The point is that the kinetic term
$\left( \p_{[\mu}B_{\nu\rho ]}\right)^2$ does not have an empty
spectrum. Thus, it can not be used as a mixing term in the master
action approach. In fact, it is  known that massless 2-forms are
equivalent to massless $D-4$ forms. In $D=3$ we could have used a
vector field to lower the order of the Maxwell theory but the dual
model would have also a Maxwell Lagrangian as kinetic term which
is equivalent on its turn to a massless scalar particle in $D=3$,
so invalidating our master action approach which crucially depends
on the specific  first-order formulation of the Maxwell theory
introduced in \cite{kmu}. In the next section we analyze the
spin-2 case.

\section{Linerarized ``New Massive Gravity'' via $\Omega$-field}

Although the minimal tensor structure to describe massive spin-2
particles is a symmetric rank-2 tensor, in order to derive a
higher derivative unitary model it is convenient, see
\cite{bfmrt}, to start with the first-order description of the
Einstein-Hilbert theory (massless action) in terms of the vielbein
and spin-connection. After linearization around a flat background
and addition of the Fierz-Pauli mass term we have:

\bea S\left\lbrack \omega,e,J \right\rbrack &=& \int \, d^D x
\left\lbrack \omega^{\mu\nu\rho}\omega_{\rho\nu\mu} -
\omega^{\mu}\omega_{\mu} + 2
\omega_{\mu\nu\alpha}K^{\alpha\mu\nu}(e) + 2
\omega_{\mu}K^{\mu}(e) \right. \nn\\ &-& \left. m^2
\left(e_{\mu\nu}e^{\nu\mu} - e^2\right) +
\omega_{\mu\nu\alpha}J^{\mu\nu\alpha} \right\rbrack \, .
\label{sfp1} \eea

\no where $e=e^{\mu}_{\,\, \mu} $ and $e_{\mu\nu}$ is a
nonsymmetric tensor which might be understood as the fluctuation
of the vielbein about a flat background while

\bea \omega_{\mu\nu\alpha} &=& - \omega_{\mu\alpha\nu} \quad ;
\quad \omega_{\alpha} = \eta^{\mu\nu}\omega_{\mu\nu\alpha} \quad ,
\label{omega} \\
K_{\alpha\mu\nu}(e) &=&  - K_{\alpha\nu\mu}(e) =
\p_{[\alpha}e_{\mu]\nu} - \p_{[\alpha}e_{\nu]\mu} +
\p_{[\nu}e_{\mu]\alpha} = \p_{\alpha} e_{[\mu\nu]} -
\p_{\mu}e_{(\nu\alpha)} + \p_{\nu}e_{\mu\alpha} \,
, \label{k3} \\
K_{\nu}(e) &=& \eta^{\alpha\mu}K_{\alpha\mu\nu}(e) = \p_{\nu} e -
\p^{\gamma}e_{\nu\gamma} \quad . \label{k0} \eea

\no The tensor $K_{\alpha\mu\nu}$ and consequently the action
$S\left\lbrack \omega,e,0 \right\rbrack $ is invariant under
linearized reparametrizations $\delta e_{\mu\nu} =
\p_{\mu}\xi_{\nu} $.  There is also a local symmetry in
(\ref{sfp1}) due to antisymmetric shifts $\delta e_{\mu\nu} =
\Lambda_{\mu\nu} = -\Lambda_{\nu\mu} \, $ , $ \, \delta
\omega_{\mu\nu\rho} = \p_{\mu}\Lambda_{\nu\rho} $.

If we integrate over the ``spin-connection'' $\omega_{\mu\nu\rho}$
in the path integral we obtain the massive Fierz-Pauli theory
${\cal L}_{FP}$ :

\bea S_{FP}\left\lbrack e,J \right\rbrack &=& \int \, d^D x
\left\lbrack {\cal L}_{LEH}(e) - m^2 \left(e_{\mu\nu}e^{\nu\mu} -
e^2\right) + K_{\mu\nu\alpha}(e)J^{\mu\nu\alpha} + {\cal L}_{JJ}
\right\rbrack \nn\\
&=& \int \, d^D x \left\lbrack {\cal L}_{FP}(e) +
K_{\mu\nu\alpha}(e)J^{\mu\nu\alpha} + {\cal L}_{JJ} \right\rbrack
\quad , \label{sfp2} \eea

\no where the linearized Einstein-Hilbert theory and the quadratic
terms in the sources are given by

\bea  {\cal L}_{LEH}(e) &=& K_{\mu}(e)K^{\mu}(e) +
K_{\mu\nu\alpha}(e)K^{\alpha\mu\nu}(e) \quad , \label{lkk} \\
{\cal L}_{JJ} &=& \frac 12 J_{\mu\nu\alpha}J^{\nu\alpha\mu} -
\frac 14 J_{\mu\nu\alpha}J^{\mu\nu\alpha} + J_{\mu}J^{\mu} \quad .
\label{ljj} \eea

\no On the other hand, if we start with (\ref{sfp1}) and integrate
over $e_{\mu\nu}$ in the path integral we derive the dual model:

\be S\left\lbrack \omega,J \right\rbrack = \int \, d^D x
\left\lbrack  {\cal L}_{\omega\omega} +
\omega^{\mu\nu\rho}w_{\rho\nu\mu} - \omega^{\mu}\omega_{\mu} +
\omega_{\mu\nu\alpha}J^{\mu\nu\alpha}
 \right\rbrack \quad .  \label{somega} \ee

\no where

\be {\cal L}_{\omega\omega} = \frac 1{m^2} \left\lbrack \left(
\p^{\nu}\omega_{\mu\alpha\nu}\right) - \frac{\left(\p \cdot \omega
\right)^2}{D-1} \right\rbrack \quad . \label{lww}\ee

\no As in the previous two sections, the kinetic term ${\cal
L}_{\omega\omega} $  has no particle content. This has been shown
in \cite{bfmrt} in the case $D=4$ via a canonical analysis in a
given gauge. Next we show that it can be generalized for arbitrary
dimensions without fixing a gauge.  Namely, we can rewrite ${\cal
L}_{\omega\omega}$ in a first-order form

\bea {\cal L}_{(1)}   &=& 2 \omega_{\mu\nu\rho}\p^{\nu}e^{\rho\mu}
+ \omega_{\mu}(\p^{\mu}e - \p_{\gamma}e^{\mu\gamma}) - m^2
\left(e_{\mu\nu}e^{\nu\mu} - e^2\right) \quad ,
\label{lww1} \\
&=& {\cal L}_{\omega\omega} - m^2 \left\lbrack
\left(e_{\mu\nu}-E_{\mu\nu}\right) \left( e^{\nu\mu} -
E^{\nu\mu}\right) - \left(e-E\right)^2 \right\rbrack \quad ,
\label{lww2} \eea

\no where

\be E_{\mu\nu} = \frac 1{m^2} \left( \p_{\mu}\omega_{\nu} +
\p^{\rho}\omega_{\mu\nu\rho} - \frac{\eta_{\mu\nu}}{D-1} \p \cdot
\omega \right) \quad . \label{E} \ee

\no After the shift $e_{\mu\nu} \to e_{\mu\nu} + E_{\mu\nu}$, it
becomes clear that ${\cal L}_{(1)}$ has the same particle content
of ${\cal L}_{\omega\omega}$. On the other hand, integrating over
$\omega_{\mu\nu\rho}$ in (\ref{lww1}) we have a functional delta
function assuring the constraint

 \be \frac{\delta
S_{(1)}}{\delta \omega_{\mu\nu\rho}} = \p^{\nu}e^{\rho\mu}-
\p^{\rho}e^{\nu\mu} + \eta^{\mu\nu}\left(\p^{\rho}e -
\p_{\gamma}e^{\rho\gamma}\right) -  \eta^{\mu\rho}\left(\p^{\nu}e
- \p_{\gamma}e^{\nu\gamma}\right) = 0 \, . \label{cons1} \ee

\no whose general solution is $e_{\mu\nu} = \p_{\mu} \phi_{\nu} $
with arbitrary $\phi_{\mu}$.
 This  can be seen from  $\eta_{\mu\nu} \frac{\delta S_{(1)}}{\delta
\omega_{\mu\nu\rho}}=0$ back in the constraint (\ref{cons1}) which
leads to $\p^{\nu}e^{\rho\mu}- \p^{\rho}e^{\nu\mu}=0$. Since $
\int \, d^D x \left(e_{\mu\nu}e^{\nu\mu} - e^2\right)$ vanishes at
$e_{\mu\nu} = \p_{\mu} \phi_{\nu} $ we conclude that ${\cal
L}_{(1)}$ and consequently ${\cal L}_{ww}$ has no particle content
in arbitrary dimensions and as such it can be used as a ``mixing
term'' in a triple master action,

\bea S_M\left\lbrack e,\omega,\tomega, J \right\rbrack &=& \int \,
d^D x \left\lbrack \omega^{\mu\nu\rho}\omega_{\rho\nu\mu} -
\omega^{\mu}\omega_{\mu} + 2
\omega_{\mu\nu\alpha}K^{\alpha\mu\nu}(e)
+ 2 \omega_{\mu}K^{\mu}(e) \right. \nn\\
&-& \left. m^2 \left(e_{\mu\nu}e^{\nu\mu} - e^2\right) -
\left({\cal L}_{\omega\omega}\right)_{\omega \to \omega + \tomega
} + \omega_{\mu\nu\alpha}J^{\mu\nu\alpha} \right\rbrack \, .
\label{sm6} \eea

\no On one hand, after the shift $\tomega_{\mu\nu\rho} \to
\tomega_{\mu\nu\rho}-\omega_{\mu\nu\rho} $ we can integrate over
$\tomega_{\mu\nu\rho}$ and $\omega_{\mu\nu\rho} $. Then, we arrive
at the massive Fierz-Pauli theory (\ref{sfp2}). On the other hand,
integrating over $e_{\mu\nu}$ followed by the integral over $
\omega_{\mu\nu\rho} $ we obtain a 4th-order model dual to the
massive Fierz-Pauli theory:

\bea S_{FP}\left\lbrack {\barf}(\tomega),J\right\rbrack &=& \int
\, d^D x \left\lbrack {\cal L}_{KK}(\barf) - m^2
\left(\barf_{\mu\nu}\barf^{\nu\mu} - \barf^2\right) +
K_{\mu\nu\alpha}(\barf)J^{\mu\nu\alpha} + {\cal L}_{JJ}
\right\rbrack \nn\\
&=& \int \, d^D x \left\lbrack {\cal L}_{FP}(\barf) +
K_{\mu\nu\alpha}(\barf)J^{\mu\nu\alpha} + {\cal L}_{JJ}
\right\rbrack \quad ,  \label{sfpf} \eea

\no Where

\be \barf_{\mu\nu} = - \frac 1{m^2} \left(
\p^{\rho}\tomega_{\mu\nu\rho} - \frac{\eta_{\mu\nu}}{D-1} \p \cdot
\tomega \right) \quad . \label{barf} \ee

  We can further simplify the dual model (\ref{sfpf}).
   Taking derivatives of the triple master action
(\ref{sm6}) with respect to the source $J^{\mu\nu\alpha}$ we
derive the strong (without contact terms) equivalence between
correlation functions:

\be \left\langle K_{\mu_1\nu_1\rho_1}[e(x_1)] \cdots
K_{\mu_N\nu_N\rho_N}[e(x_N)]\right\rangle_{S_{FP}(e)} =
\left\langle K_{\mu_1\nu_1\rho_1}[\barf(x_1)] \cdots
K_{\mu_N\nu_N\rho_N}[\barf(x_N)]\right\rangle_{S_{FP}[\barf(\tomega)]}
\, . \label{cf2} \ee

We infer the local dual map
$K_{\mu\nu\rho}[e(x)]=K_{\mu\nu\rho}[\barf (x)]$ which is
equivalent to $K_{\mu\nu\rho}[e-\barf ]=0$ whose general solution
is pure gauge $\left(e- f\right)_{\mu\nu} = \p_{\mu}\xi_{\nu} $.
So we can write down

\be e_{\mu\nu} = \barf_{\mu\nu}  + \p_{\mu}\xi_{\nu} = - \frac
1{m^2} \left( \p^{\rho}\tomega_{\mu\nu\rho} -
\frac{\eta_{\mu\nu}}{D-1} \p \cdot \tomega \right) +
\p_{\mu}\xi_{\nu} = \p^{\rho}\Omega_{\mu\nu\rho} +
\p_{\mu}\tilde{\xi}_{\nu} \, , \label{dmap1} \ee

\no where

\bea \Omega_{\mu\nu\rho} &=& -\frac 1{m^2}\left(
\tomega_{\mu\nu\rho} + \frac{\eta_{\mu\rho}\tomega_{\nu} -
\eta_{\mu\nu}\tomega_{\rho}}{D-1} \right)  \quad , \label{omegat}
\\
\tilde{\xi} &=& \xi_{\nu} + \frac{\tomega_{\nu}}{m^2(D-1)}  \quad
. \label{xit} \eea

\no Thus, we deduce the local dual map between the nonsymmetric
tensor $e_{\mu\nu}$  of the Fierz-Pauli  theory and the traceless
($\eta^{\mu\nu}\Omega_{\mu\nu\rho}=\Omega_{\rho}=0$) field
$\Omega_{\mu\nu\rho}$:

\be e_{\mu\nu} = f_{\mu\nu} + \p_{\mu}\tilde{\xi}_{\nu} \quad ,
\label{dmap2} \ee

\no with

\bea f_{\mu\nu} &=& \p^{\rho}\Omega_{\mu\nu\rho} \quad ,
\label{fmn}
\\
\eta^{\mu\nu}f_{\mu\nu} &=& f  =0 \quad, \label{f} \\
\p^{\nu}f_{\mu\nu} &=& 0 \quad . \label{fm} \eea

Due to the subsidiary conditions (\ref{f}) and (\ref{fm}), when we
substitute $e_{\mu\nu} = f_{\mu\nu} + \p_{\mu}\tilde{\xi}_{\nu}$
in the massive Fierz-Pauli theory (\ref{sfp2}), the arbitrary
functions $\tilde{\xi}_{\nu}$ drop out and we recover the
4th-order dual model (\ref{sfpf}) with $\barf_{\mu\nu}$ replaced
by $f_{\mu\nu}$. Dropping the source terms, the linearized ``New
Massive Gravity'' Lagrangian in arbitrary dimensions can be
written  as\footnote{Since the first two terms of (\ref{lnmg})
only depend upon $\Omega_{(\mu\nu)\rho} $, it is tempting to split
$\Omega_{\mu\nu\rho} = \Omega_{[\mu\nu]\rho} +
\Omega_{(\mu\nu)\rho} $ and drop $\Omega_{[\mu\nu]\rho}$. However,
the corresponding theory in terms of $\Omega_{(\mu\nu)\rho} $
contains ghosts. The point is that $\Omega_{[\mu\nu]\rho}$ and
$\Omega_{(\mu\nu)\rho} $ are not independent variables due to the
property $\Omega_{\mu\nu\rho} = - \Omega_{\mu\rho\nu}$. For
instance, $\Omega_{(13)1}=\Omega_{[13]1}$.}

\bea {\cal L}_{\Omega} &=& \frac 12
\left(\p^{\mu}f_{\mu\nu}\right) \left(\p^{\alpha}f_{\alpha}^{\quad
\nu} \right) - \p^{\nu}f^{(\alpha\beta)}\p_{\nu}f_{(\alpha\beta)}
- m^2 f_{\mu\nu}f^{\nu\mu} \quad , \label{lnmgf}\\
&=& \frac 12 \left(\p^{\mu}\p^{\rho}\Omega_{\mu\nu\rho}\right)^2 -
\left\lbrack
\p^{\nu}\p^{\rho}\Omega_{(\alpha\beta)\rho}\right\rbrack^2 - m^2
\left(\p^{\rho}\Omega_{\mu\nu\rho}\right)\left(\p^{\alpha}\Omega^{\nu\mu}_{\quad\alpha}\right)
\, . \label{lnmg} \eea

Although our triple master action approach already  proves that
${\cal L}_{\Omega}$ only contains one massive ``spin-2'' particle
in the spectrum, it is instructive to check it directly from
${\cal L}_{\Omega}$ as follows.

First, we note that the equations of motion of ${\cal L}_{\Omega}$
can be compactly written as a set of zero ``curvature''
conditions:

\be \p_{\rho}T_{\beta\alpha} - \p_{\beta}T_{\rho\alpha} = 0 \quad
. \label{eqm4} \ee

\no where

\be T_{\mu\nu} = \Box f_{(\mu\nu)} - m^2 \, f_{\mu\nu} - \frac 12
\p_{\nu}\p^{\beta}f_{\beta\mu} \quad , \label{tmn}\ee

\be  T = 0 = \p^{\nu} T_{\mu\nu} \quad , \label{sc} \ee

\no with $f_{\mu\nu}$ given in (\ref{fmn}). The simplicity of
(\ref{eqm4}) is connected with the fact that $S_{\Omega}=\int d^D
x \, {\cal L}_{\Omega}$ depends upon the traceless tensor
$\Omega_{\mu\nu\rho}$ only through
$f_{\mu\nu}=\p^{\rho}\Omega_{\mu\nu\rho}$. One can deduce
(\ref{eqm4}) either directly from (\ref{lnmg}) or via the
functional chain rule:

\be \frac{\delta S_{\Omega}}{\delta \Omega_{\mu\nu\rho}(x)} = \int
d^D z\, \frac{\delta S_{\Omega}}{\delta
f_{\alpha\beta}(z)}\frac{\delta f_{\alpha\beta}(z)}{\delta
\Omega_{\mu\nu\rho}(x)} = - \frac 12 \left( \p^{\rho} \frac{\delta
S_{\Omega}}{\delta f_{\mu\nu}(x)} - \p^{\nu} \frac{\delta
S_{\Omega}}{\delta f_{\mu\rho}(x)}\right) \, . \label{chain} \ee

\no In order to prove (\ref{chain}) we have used the properties
$\Omega_{\rho} = 0$ , $ \Omega_{\mu\nu\rho} = -
\Omega_{\mu\rho\nu}$ and the identities $\eta^{\mu\nu}\left(\delta
S_{\Omega}/\delta f_{\mu\nu}\right)=0$ ,  $\p_{\mu}\left(\delta
S_{\Omega}/\delta f_{\mu\nu}\right)=0$.

The general solution to (\ref{eqm4}) consistent with (\ref{sc}) is
given in terms of an arbitrary transverse vector,

\be T_{\mu\nu} = \Box f_{(\mu\nu)} - m^2 \, f_{\mu\nu} - \frac 12
\p_{\nu}\p^{\beta}f_{\beta\mu} = \p_{\mu}A_{\nu}^T \quad , \quad
\p^{\mu}A_{\mu}^T =0 \, , \label{seqm4} \ee

\no In order to solve (\ref{seqm4}) we first notice that
$S_{\Omega} $ is invariant under the gauge transformations:

\be \delta\Omega_{\mu\nu\rho} = \p_{\mu}B_{\nu\rho} +
\frac{\eta_{\mu\rho}\p^{\alpha}B_{\alpha\nu}-\eta_{\mu\nu}\p^{\alpha}B_{\alpha\rho}}{D-1}
+ \p^{\alpha}\Lambda_{[\alpha\nu\rho]\mu} \quad , \label{gt2} \ee

\no where $B_{\mu\nu} = - B_{\nu\mu}$ and
$\p^{\alpha}\Lambda_{[\alpha\mu\rho]}^{\quad\quad \mu} =0$. After
a trivial redefinition of $B_{\mu\nu}$ we have $\delta f_{\mu\nu}
= \p_{\mu}\left( \p^{\alpha}B_{\alpha\nu} \right) $. Consistency
with (\ref{seqm4}) requires :

\be \delta A_{\mu}^T = \left(\frac{\Box}2 - m^2 \right)
\p^{\nu}B_{\nu\mu} \quad . \label{gta} \ee.

\no In order to find a convenient gauge condition we notice from $
\delta f_{\mu\alpha}=\p_{\mu}\left( \p^{\nu}B_{\nu\alpha} \right)
$ and from $\p^{\mu}$ applied on (\ref{seqm4}), respectively, that

\be \delta \p^{\mu}f_{\mu\alpha}= \Box \left(
\p^{\nu}B_{\nu\alpha} \right) \equiv \Box U_{\alpha}^T \quad ;
\quad \p^{\mu}f_{\mu\alpha} = \frac{\Box}{m^2}
\left(\frac{\p^{\mu}f_{\mu\alpha}}2 - A_{\alpha}^T\right) \equiv
\Box V_{\alpha}^T \quad , \label{eqm5} \ee

\no From (\ref{eqm5}) it is natural to fix the gauge:

\be \p^{\mu}f_{\mu\alpha} = 0 \quad . \label{gauge} \ee

\no Back in the second equation of (\ref{eqm5}) we have $\Box
A_{\mu}^T =0$. All the equations so far are invariant under
residual harmonic gauge transformations: $\Box
\left(\p^{\mu}B_{\mu\alpha} \right)=0$.  From (\ref{gta}) we have
$\delta A_{\alpha}^T = - m^2 \left(\p^{\mu}B_{\mu\alpha} \right) =
-m^2 U_{\alpha}^T $. So we can use the residual symmetry to set
$A_{\alpha}^T =0$. The first equation (\ref{seqm4}) becomes $ \Box
f_{(\nu\mu)} - m^2 f_{\mu\nu} = 0 $ which leads to

\be f_{[\mu\nu]} =0 \quad , \quad \left(\Box - m^2
\right)f_{\mu\nu} =0 \quad . \label{fpc} \ee

\no The equations (\ref{fpc}) and the identities $f=0=
\p^{\mu}f_{\alpha\mu}$ correspond to the Fierz-Pauli conditions.
Since $f_{\mu\nu} = \p^{\rho}\Omega_{\mu\nu\rho} $ is invariant
under the unfixed gauge transformations $\delta
\Omega_{\mu\nu\rho} = \p^{\alpha}\Lambda_{[\alpha\nu\rho]\mu}$
where $\p^{\alpha}\Lambda_{[\alpha\mu\rho]}^{\quad\quad \mu} =0$,
the particle content of $S_{\Omega}$ corresponds indeed to one
massive ``spin-2'' particle as expected.

\section{Recovering ``New Massive'' 3D and 4D Gravity}

In order to make contact with the known theories of ``New Massive
Gravity'' in $D=2+1$ \cite{bht} and in $D=3+1$ \cite{bfmrt} we
remark that the traceless $\Omega$-field has $D(D+1)(D-2)/2$
independent components. In $D=3$ we have $6$ degrees of freedom
which coincides with a symmetric rank-2 tensor. Inspired by
\cite{bfmrt} we use the Levi-Civita symbol and define the linear
change of variables

\be \Omega_{\mu\nu\rho} =
\epsilon_{\nu\rho\alpha}h^{\alpha}_{\quad \mu} \quad \to \quad
f_{\mu\nu} = \p^{\rho}\Omega_{\mu\nu\rho} = -
\hat{E}_{\nu}^{\quad\alpha}  h_{\alpha\mu} \label{mapd3} \ee

\no where $\hat{E}_{\nu\alpha} =
\epsilon_{\nu\alpha\gamma}\p^{\gamma}$. Due to the traceless
condition $\eta^{\mu\nu}\Omega_{\mu\nu\rho}=0$ we must have
$h_{\mu\nu}=h_{\nu\mu}$. By replacing $ f_{\mu\nu} = -
\hat{E}_{\nu}^{\quad\alpha} h_{\alpha\mu}$ in (\ref{lnmg}) and
using the identity $\hat{E}_{\mu\alpha}\hat{E}_{\beta\nu} = \Box
\left( \theta_{\mu\nu}\theta_{\alpha\beta} -
\theta_{\mu\beta}\theta_{\alpha\nu}\right)$ we arrive (up to an
overall factor $2\, m^2$ ) at the linearized version of the new
massive gravity of \cite{bht}:

\be {\cal L}_{NMG}^{3D}= \frac 12 h_{\mu\beta} \Box^2 \left\lbrack
2\theta^{\mu\nu}\theta^{\alpha\beta} - \theta^{ \mu\beta}\theta^{
\alpha\nu} \right\rbrack h_{\nu\alpha} - m^2
h_{\mu\beta}\hat{E}^{\mu\alpha}\hat{E}^{\beta\nu}h_{\alpha\nu} \,
. \label{nmg3} \ee

 The linearized reparametrization invariance $\delta h_{\mu\nu} =
\p_{\mu}\xi_{\nu} + \p_{\nu}\xi_{\mu} $ of $S_{NMG}^{3D} $ becomes
$\delta f_{\mu\nu} = \p_{\mu} \Phi_{\nu}^T$ where $\Phi_{\nu}^T =
-\hat{E}_{\nu\beta}\xi^{\beta}$ which is  of course a symmetry of
(\ref{lnmg}) as a consequence of the $B_{\mu\nu}$ gauge invariance
(\ref{gt2}). Notice also that in $D=2+1$ the $\Lambda$-symmetry,
see (\ref{gt2}), is a subset of the $B_{\mu\nu}$-symmetry.

In $D=4$ the analogue of the map (\ref{mapd3}) is given by

\be \Omega_{\mu\nu\rho} = \epsilon_{\nu\rho}^{\quad
\gamma\delta}T_{[\gamma\delta]\mu} \quad \to \quad f_{\mu\nu} =
\p^{\rho}\Omega_{\mu\nu\rho} =
\hat{E}_{\nu}^{\,\,\gamma\delta}T_{[\gamma\delta]\mu}
\label{mapd4} \ee

\no where $\hat{E}_{\mu\nu\alpha} =
\epsilon_{\mu\nu\alpha\gamma}\p^{\gamma}$. The traceless condition
$\Omega_{\mu}=0$ requires $T_{[[\gamma\delta]\mu]}=0$. The
$\Omega$-field has 20 components in $D=4$  which is the same
number of components of a tensor $T_{[\gamma\delta]\mu}$ without
totally antisymmetric part. If we replace $\Omega_{\mu\nu\rho} =
\epsilon_{\nu\rho}^{\quad \gamma\delta}T_{[\gamma\delta]\mu}$ in
(\ref{lnmg}) we have

\be {\cal L} = T_{[\mu\nu]\beta} \left( \frac{\Box}2 - m^2 \right)
G^{[\mu\nu]\beta}(T) - \frac 14 T_{[\mu\nu]\alpha}\Box
\theta^{\alpha\delta}\hat{E}^{\mu\nu}_{\quad
\beta}E^{\lambda\epsilon\beta}T_{[\lambda\epsilon]\delta} \, .
\label{lnmg4a} \ee

\no where, see \cite{bfmrt}, $G^{[\mu\nu]\rho}(T) = -
\hat{E}^{[\mu\nu]\beta}\hat{E}^{[\rho\delta]\epsilon}T_{[\delta\epsilon]\beta}/2
$. In order to bring (\ref{lnmg4a}) to the form given in
\cite{bfmrt} we have to rearrange its last term. We have found
convenient to use the 4D identity
$\epsilon_{[\mu\nu\beta\gamma}a_{\delta ]}=0$ twice. First with
$a_{\delta} \to \p_{\delta}$ and secondly with  $a_{\delta} \to
\eta_{\delta\alpha} $ and multiplying the result with
$\p^{\gamma}$. Back in (\ref{lnmg4a}) we get the linearized  ``New
Massive'' 4D gravity (up to an overall factor $2\, m^2$ ) as given
in formula (29) of \cite{bfmrt}:

\be {\cal L}_{LNMG}^{4D} = T_{[\mu\nu]\beta}\left(\Box -m^2
\right) G^{[\mu\nu]\beta}(T) + T_{[\mu\nu]\beta}\, \Box
\theta^{\beta [ \mu}G^{\nu ]}(T) \quad , \label{lnmg4} \ee

\no where $ G_{\mu}(T) = \eta^{\nu\rho}G_{[\mu\nu]\rho}(T)$. It
can be shown that the local symmetries of (\ref{lnmg4}) mentioned
in \cite{bfmrt} are equivalent to the gauge symmetries
(\ref{gt2}).

For arbitrary dimensions we can introduce a rank-(D-1) tensor and
derive dual descriptions via $\Omega_{\mu\nu\rho} =
\epsilon_{\nu\rho}^{\quad \alpha_1 \cdots
\alpha_{D-2}}T_{[\alpha_1 \cdots \alpha_{D-2}]\mu} $. Due to the
traceless condition we must have $T_{[\alpha_1 \cdots
\alpha_{D-2}\mu ]}=0 $. However, for $D \ge 5$ we better stick,
for simplicity, to the rank-3 description given by ${\cal
L}_{\Omega}$.

Last, we comment on the linearized Weyl symmetry of 4th-order
terms of $S_{\Omega}$ which is specially relevant for a possible
nonlinear completion of $S_{\Omega}$ and its renormalizability.

The linearized Weyl symmetry of the 4th-order terms of
$S_{NMG}^{3D} $ and $S_{NMG}^{4D} $ correspond respectively to the
transformations  $\delta_w h_{\mu\nu} = \eta_{\mu\nu}\phi$ and
$\delta_w T_{[\alpha\beta]\mu} =
\left(\eta_{\alpha\mu}\phi_{\beta} -
\eta_{\beta\mu}\phi_{\alpha}\right)/2 $. In terms of the
$\Omega$-field we have $\delta_w
\Omega_{\mu\nu\rho}=\epsilon_{\mu\nu\rho}\phi $ and $\delta_w
\Omega_{\mu\nu\rho}=\epsilon_{\mu\nu\rho\beta}\phi^{\beta} $. They
can be easily generalized to arbitrary dimensions:

\be \delta_w\Omega_{\mu\nu\rho}=\epsilon_{\mu\nu\rho\alpha_1
\cdots \alpha_{D-3}}\phi^{\alpha_1 \cdots \alpha_{D-3}} \quad \to
\quad \delta_w f_{\mu\nu} = - (-)^D \hat{E}_{\mu\nu\alpha_1 \cdots
\alpha_{D-3}} \phi^{\alpha_1 \cdots \alpha_{D-3}} \, .
\label{weyl} \ee

\no where $\hat{E}_{\mu\nu\alpha_1 \cdots
\alpha_{D-3}}=\epsilon_{\mu\nu\alpha_1 \cdots
\alpha_{D-3}\gamma}\p^{\gamma}$.

 Since the 4th-order terms of
(\ref{lnmgf}) depend upon $\p^{\mu}f_{\mu\nu}$ and $f_{(\mu\nu)}$
it is clear they are invariant under the Weyl transformations
(\ref{weyl}) which are broken by the 2nd-order mass term. Thus, in
arbitrary dimensions, a possible nonlinear completion of the ``New
Massive Gravity'' ${\cal L}_{\Omega}$ may have renormalizability
problems, see \cite{mo} for the $3D$ NMG case. Since part of the
degrees of freedom of the $\Omega$-field will not be present in
the 4th-order terms, their UV behavior will be ruled by the
2nd-order mass term $\sim 1/p^2$ which is a problem for
renormalizability  unless the nonlinear (self-interacting) terms
are also Weyl invariant. This is not the case of the $NMG$ in
$D=3$. See comments in a similar vein in \cite{deserprl,deg}.

\section{Conclusion and Outlook}

Here we have shown that a triple master action approach can be
used to deduce a higher-rank 4th-order (in derivatives)
description of spin-0, spin-1 and spin-2 massive particles in
arbitrary dimensions. By adding sources we have explained why the
dual higher-derivative  theories can be obtained via a strong
local map which contains one derivative of a higher rank field and
works directly at action level. This is to be contrasted with
other examples in the literature where local maps between theories
of different number of derivatives only hold at the level of
equations of motion like, e.g., the duality in $D=3$  between the
spin-1 self-dual model \cite{pnt} and the Maxwell-Chern-Simons
theory where the map $f_{\mu} =
\epsilon_{\mu\nu\beta}\p^{\nu}A^{\beta}/m$ works on shell but
leads to ghosts if used at action level (off-shell). The key point
for the off-shell usefulness of the dual map is the absence of
contact terms in the equivalence between correlation functions.

In the spin-2 case we have shown that the replacement
$e_{\mu\nu}=\p^{\rho}\Omega_{\mu\nu\rho}$  in the massive
Fierz-Pauli action with a nonsymmetric rank-2 tensor  leads
directly to the unitary 4th-order model ${\cal L}_{\Omega}$, see
(\ref{lnmg}), which might be interpreted as a generalization of
the linearized ``New Massive'' 3D and 4D gravity to arbitrary
dimensions. We have shown that ${\cal L}_{\Omega}$ indeed
describes a massive ``spin-2'' particle in arbitrary dimensions
and reduces to the known ``New Massive'' gravity theories in $D=3$
and $D=4$. We have pointed out the importance of the traceless
condition $\eta^{\mu\nu}\Omega_{\mu\nu\rho}=0$ in order to
simplify the dual 4th-order model and reduce the number of initial
degrees of freedom. The equations of motion can be written in a
simple form as zero ``curvature'' conditions.

Since the structure of the theory in $D$-dimensions is basically
the same of the $D=3$ case, i.e., one second-order term plus a
``Weyl'' invariant 4th-order term, it is expected that a possible
nonlinear completion of ${\cal L}_{\Omega}$ might suffer from the
same renormalizability problems of the $D=3$ case, see \cite{mo}
and comments in \cite{deserprl,deg}, worsened by the higher
dimensionality.

 Even if we are able to add Weyl invariant
interactions to ${\cal L}_{\Omega}$, this will probably require
the vertices to depend upon
$f_{(\mu\nu)}=\p^{\rho}\Omega_{(\mu\nu)\rho}$ which contains
already one derivative such that the net gain in the power
counting is zero. The same happens already in the spin-0 case
treated in the second section. If we plug $\phi = \p \cdot B $ in
nonlinear terms in the scalar field, the extra $1/p^2$ factor in
the UV behavior of the vector field propagator will be canceled by
the extra $p^2$ factor on each internal line of Feyman diagrams
due to the derivative vertices. So we better look for alternative
higher-derivative theories where all propagating degrees of
freedom are present in the highest derivative term.

It may be useful to comment that if we make a weak field expansion
$g_{\mu\nu}= \eta_{\mu\nu} + h_{\mu\nu}$ in the nonlinear ``New
Massive'' gravity of \cite{bht} in $D=3$ and  replace
$h_{\mu\nu}=h_{\mu\nu}(\Omega)$ where $h_{\mu\nu}(\Omega)$ is
obtained by inverting the map
$\Omega_{\mu\nu\rho}=\epsilon_{\nu\rho\beta}h^{\beta}_{\,\,\mu}$.
we obtain a consistent ghost-free (beyond tree level, see
\cite{rgpty}) interacting theory for the mixed symmetry
$\Omega$-field. Such formulation may be eventually relevant for
the addition of interactions to ${\cal L}_{\Omega}$ for $D>3$.
Another possible approach is to look for a nonlinear completion of
the gauge transformations (\ref{gt2}) altogether with consistent
(gauge invariant) vertices to be added to ${\cal L}_{\Omega}$.

A further direction to follow is the generalization of the triple
master action approach to higher-spin theories.

Finally, while typing this work we became aware of \cite{jm} where
the issue of a linearized $D$-dimensional ``New Massive Gravity''
has also been addressed.

\section{Acknowledgements}

The work of D.D. is partially supported by CNPq. R.C.S thanks
CAPES for a Master-fellowship during his stay at UNESP-Campus de
Guaratinguet\'a where this work was carried out.

\end{document}